\documentclass[aps,preprint,showpacs,preprintnumbers,amsmath,amssymb]{revtex4}
\usepackage[dvips]{graphicx}
\usepackage{float}
\begin{document}

\title{Fermion production by a dependent of time electric field in de Sitter universe}
\author{ Cosmin Crucean \thanks{E-mail:~~crucean@physics.uvt.ro}\\
{\small \it West University of Timi\c soara,}\\
{\small \it V. Parvan Ave. 4 RO-300223 Timi\c soara,  Romania}}

\begin{abstract}
Fermion production by the electric field of a charge on de Sitter
expanding universe is analyzed. The amplitude and probability of
pair production are computed. We obtain from
our calculations that the modulus of the momentum is no longer
conserved and that there are probabilities for production
processes where the helicity is no longer conserved. The rate of
pair production in an electric field is found to be important in the
early universe when the expansion factor was large comparatively
with the particle mass.
\end{abstract}

\pacs{04.62.+v}

\maketitle

\section{Introduction}
Pair production by electric field is known as the Schwinger
 effect \cite{rf:25}, and even if this phenomenon was not yet observed,
it might be possible to be experimentally proven in the next
years. The method employed in obtaining this result is based on
the non-perturbative structure of the QED vacuum physics, when
strong fields are considered. However, pair production by a weak electric field
is in principle possible if this field is coupled with strong gravity. This kind of conditions
were achieved only in the early universe. In the present paper we want to
prove that pair production by the electric field of a charge could have arisen in the early universe due to
the large values of the expansion factor.

The problem of fermion production by a constant electric field on an
expanding universe was the subject of a few investigations
\cite{19,20,27,31} . The main results of the mentioned papers
show that there are indeed nonvanishing rates for pair production
in an electric field. However, pair
production in the early universe was not addressed in \cite{19,20,27} due to some
constrains which are imposed to the particle mass, which must be
larger than the expansion factor of the space.

It is important to specify that in
our calculations we will use the electric field
produced by a point charge (or a distribution of charges) in the de Sitter universe, which is a time dependent
 quantity. To the best of our knowledge, this kind of electric field was
not considered in the investigations related to the pair production
in an expanding de Sitter universe. The work of L.
Parker \cite{8,9,10} shows that we must expect a considerable amount
of pair production from vacuum in the early universe.
Then we must expect the phenomenon of pair production to be amplified
in the presence of an external electric field coupled with gravity.

In the present paper, we want to prove that
the electric field produced by a charge, or a distribution of
charges can produce large amounts of fermions in the early
expansion conditions.
Our method is based on perturbations and was used in a number of
investigations for calculating QED processes in de Sitter
universe \cite{16,17,18} . To the best of our knowledge, the method based on
perturbations was not used to calculate the probability of pair
production in electric fields on de Sitter space. We will compute
the amplitude corresponding to the process $vacuum \rightarrow
e^-+e^+$ in the presence of an electric field. The time reversed
process can be also calculated, and represents the process in
which the pair is absorbed by the electric field. As we know, a
point charge can produce an electric field with intensity
decreasing with the square of distance to the source (charge). In flat space,
this kind of electric field doesn't produce pairs of fermions.
However, in the early universe, this phenomenon can not be excluded.

The second section begins with a short review of the free fields
theory. In the third section we compute the amplitude and
probability of pair production in an electric field and we analyze
the physical consequences of our calculations. The fourth section is
dedicated to the calculation of the total number of produced fermions and in section
five we present our conclusions.

\section{Preliminaries}

The theory of the free Dirac field on de Sitter space is well known
and in the Appendix is given only the form of the fundamental
solutions, which have a definite momentum and helicity. These
solutions will help us calculate the form of our amplitude.

We will focus in this section in establishing the form of the
electric field in the de Sitter metric \cite{3} :
\begin{equation}\label{metr}
ds^{2}=dt^{2}-e^{2\omega t}d\vec{x}^{2}=\frac{1}{(\omega
t_{c})^2}(dt_{c}^{2}-d\vec{x}^{2}),
\end{equation}
where $\omega$ is the expansion factor $(\omega>0)$ and the
conformal time is defined as $t_{c}=-\frac{e^{-\omega
t}}{\omega}$. The form of the line element allows us to choose the
simple Cartesian gauge with the nonvanishing tetrad components
\begin{equation}
e^{0}_{\widehat{0}}=e^{-\omega t}
;e^{i}_{\widehat{j}}=\delta^{i}_{\widehat{j}}\,e^{-\omega
t}\,;e^{\widehat{0}}_{0}=e^{\omega
t}\,;e^{\widehat{i}}_{j}=\delta^{\widehat{i}}_{j}\,e^{\omega t}
\end{equation}
so that $e_{\widehat{\mu}}=e^{\nu}_{\widehat{\mu}}e_{\nu}$ and
have the orthonormalization properties
$e_{\widehat{\mu}}e_{\widehat{\nu}}=\eta_{\widehat{\mu}\widehat{\nu}},\\
  \widehat{e}^{\widehat{\mu}}e_{\widehat{\nu}}=\delta^{\widehat{\mu}}_{\widehat{\nu}}$,
with respect to the Minkowski metric $\eta=diag(1,-1,-1,-1)$.

The theory of the free electromagnetic field on de Sitter space is simple,
if we recall that the de Sitter metric is conformal with the
Minkowski metric. At a conformal transformation the metric tensor
$g_{\alpha \beta}$ can always be expressed in the form $g_{\alpha
\beta}=\Omega\,\eta_{\alpha \beta}\,,g^{\alpha
\beta}=\Omega^{-1}\,\eta^{\alpha \beta}$, where $\eta_{\alpha
\beta}$ is the Minkowski metric and $\Omega=\frac{1}{(\omega
t_{c})^2}$. Using this transformation, it was shown that the four
vector potential in de Sitter space can be expressed in terms of
the corresponding four vector potential on Minkowski space as
follows \cite{24} :
\begin{equation}\label{conform}
 A_{\mu\,S}=A_{\mu\,M}\,\,\,, A_{S}^{\mu}=\Omega^{-1}\, A_{M}^{\mu}.
\end{equation}
The electric field of a charge will be under the expansion conditions
a time-dependent quantity. This can be seen from the expression of the electromagnetic field tensor in the de Sitter metric. The electromagnetic field tensor in
de Sitter space can be expressed in terms of the electromagnetic field tensor from
Minkowski space, in order to preserve the conformal invariance:
\begin{equation}\label{m}
F_{\mu\nu\,S}=F_{\mu\nu\,M}\,\,\,,F^{\mu\nu}_{S}=\Omega^{-2}F^{\mu\nu}_{M},
\end{equation}
Then in a local Minkowski frame, equation (\ref{m}) becomes,
$F^{\hat\mu\hat\nu}_{S}=e^{\hat\mu}_{\alpha}e^{\hat\nu}_{\beta}F^{\alpha\beta}_{S}=e^{\hat\mu}_{\alpha}e^{\hat\nu}_{\beta}\Omega^{-2}F^{\alpha\beta}_{M}$.
Now, if we consider the electric field produced by a charge $Q$, or
a distribution of charges $\sum_{i}Q_{i}$, we will find in de
Sitter space, that at distance $|\vec{x}|$ from the charge, the
intensity of the electric field is:
\begin{equation}
\vec{E}=-\frac{\partial\vec{A}}{\partial
t}=\frac{Q}{|\vec{x}|\,^2}e^{-2\omega t}\vec{n},
\end{equation}
where $\vec{n}$ is the unit vector which gives the direction and
orientation of the intensity of electric field. It is now simple
to establish the form of the vector potential that will be
associated to this electric field:
\begin{equation}\label{e1}
\vec{A}=\frac{Q}{2\omega|\vec{x}|\,^2}e^{-2\omega t}\vec{n}.
\end{equation}
This expression will be further used to calculate the amplitude of pair production by an
electric field in de Sitter universe.

\section{Transition amplitude and probability}

In this section we will calculate the transition amplitude and
probability, for pair production in an electric field and we will
explore the physical consequences of our result. The form of the
transition amplitude can be established using the same methods as
in the flat space case, as was shown in \cite{12,16,17,18,21,22} .
For pair production in an external electromagnetic field,
supposing that the fields are coupled by the elementary electric
charge $e$, the expression of the transition amplitude is:
\begin{equation}\label{ampl}
\mathcal{A}_{e^-e^+}=-ie \int d^{4}x \left[-g(x)\right]^{1/2}\bar
U_{\vec{p},\,\lambda}(x)\vec{\gamma}\cdot\hat{\vec{A}}(x)
V_{\vec{p}\,\,',\,\lambda'}(x)
\end{equation}
where $e$ is the unit charge of the field and the fields
$U_{\vec{p},\,\lambda}(x)\,\,,V_{\vec{p}\,\,',\,\lambda'}(x)$, are
supposed to be exact solutions of the free Dirac equation in the
momentum basis \cite{4} . Using the fundamental solutions of the Dirac
equation given in equation (\ref{sol}) from Appendix and the potential (\ref{e1}) and setting $Q=e$, we finally arrive at the amplitude:
\begin{eqnarray}\label{in}
\mathcal{A}_{e^-e^+}=-i \frac{e^{2}
\sqrt{pp\,'}}{32|\vec{p}+\vec{p}\,'|}\left[-sgn(\lambda\lambda\,')e^{-\pi
k}\int_0^{\infty} dz
z^2H^{(2)}_{\nu_{-}}(p z)H^{(2)}_{\nu_{-}}(p\,'z)\right.\nonumber\\
\left.+e^{\pi k}\int_0^{\infty} dz z^2 H^{(2)}_{\nu_{+}}(p
z)H^{(2)}_{\nu_{+}}(p\,'
z)\right]\xi^{+}_{\lambda}(\vec{p}\,)(\vec{\sigma}\cdot\vec{n})\eta_{\lambda'}(\vec{p}\,\,'),
\end{eqnarray}
The result of the spatial integral \cite{7}
was included in the above amplitude. In the temporal integral a new variable $z=-t_{c}$ was introduced and $sgn$ is the signum
function. To solve these integrals we use the relation between
Hankel functions and Macdonald functions \cite{11,12,23} , arriving
in this way at the integrals discussed in Appendix. The result of
our calculations can be expressed in terms of unit step functions $\theta$, Euler gamma
functions $\Gamma$ and Gauss hypergeometric functions $_{2}F_{1}$:

\begin{eqnarray}\label{final}
\mathcal{A}_{e^-e^+}= \frac{e^{2}
}{64\pi|\vec{p}+\vec{p}\,'|}\left[p^{-2}\theta(p-p\,')f^*_{k}\left(\frac{p\,'}{p}\right)+
p\,'^{-2}\theta(p\,'-p)f^*_{k}\left(\frac{p}{p\,'}\right)\right.\nonumber\\
\left.-sgn(\lambda\lambda\,')\left(p^{-2}\theta(p-p\,')f_{k}\left(\frac{p\,'}{p}\right)+
p\,'^{-2}\theta(p\,'-p)f_{k}\left(\frac{p}{p\,'}\right)\right)\right]\xi^{+}_{\lambda}(\vec{p}\,)(\vec{\sigma}\cdot\vec{n})\eta_{\lambda'}(\vec{p}\,\,'),
\nonumber\\
\end{eqnarray}
where the functions
$f_{k}\left(\frac{p}{p\,'}\right)$, which define the amplitude are given by:
\begin{eqnarray}\label{fc}
f_{k}\left(\frac{p}{p\,'}\right)=
\left(\frac{p}{p\,'}\right)^{1-ik}\Gamma\left(2-ik\right)\Gamma\left(1+ik\right)
\,_{2}F_{1}\left(\frac{3}{2},2-ik;3;1-\left(\frac{p}{p\,'}\right)^{2}\right)\nonumber\\
=\frac{4}{\sqrt{\pi}}\left(\frac{p}{p\,'}\right)^{1-ik}\Gamma\left(ik-\frac{1}{2}\right)\Gamma\left(2-ik\right)\,_{2}F_{1}\left(\frac{3}{2},2-i
k;\frac{3}{2}-ik;\left(\frac{p}{p\,'}\right)^{2}\right)\nonumber\\
+\frac{4}{\sqrt{\pi}}\left(\frac{p}{p\,'}\right)^{ik}\Gamma\left(\frac{1}{2}-ik\right)\Gamma\left(1+ik\right)\,_{2}F_{1}\left(\frac{3}{2},1+i
k;\frac{1}{2}+ik;\left(\frac{p}{p\,'}\right)^{2}\right),
\end{eqnarray}
where he second equality is obtained when we use (\ref{a1}) and the function $f_{k}\left(\frac{p\, '}{p}\right)$ is obtained when one makes
the substitution $p\rightleftarrows p\,'$. The parameter $k=m/\omega$ is the
particle mass/expansion factor ratio.

We will explore in the rest of the paper the physical
consequences of our calculations. It is obvious from equation
(\ref{final}), that the only possible transitions are those in
which the momenta of the electron and positron are not equal. From
here we conclude that the law of conservation for the modulus of
momentum is lost in this process. The background which enables the
breaking of momentum conservation law is the external electric
field, not the de Sitter geometry, because the geometry
(\ref{metr}) has spatial translation invariance as an exact
symmetry, so the momentum is conserved \cite{26} .

Further let us study the properties of our amplitude by drawing
the graphs of the real and imaginary parts of
$f_{k}\left(\frac{p}{p\,'}\right)$ (the analysis is similar for
function  $f_{k}\left(\frac{p\,'}{p}\right)$), as function of
parameter $k=m/\omega$,  for different values of the ratio
$\frac{p}{p\,'}=\chi(=\frac{p\,'}{p})$. The parameter $k$ encodes the influence of
space expansion upon the pair production process.

\begin{figure}
\centerline{\includegraphics[width=8 cm,height=6 cm]{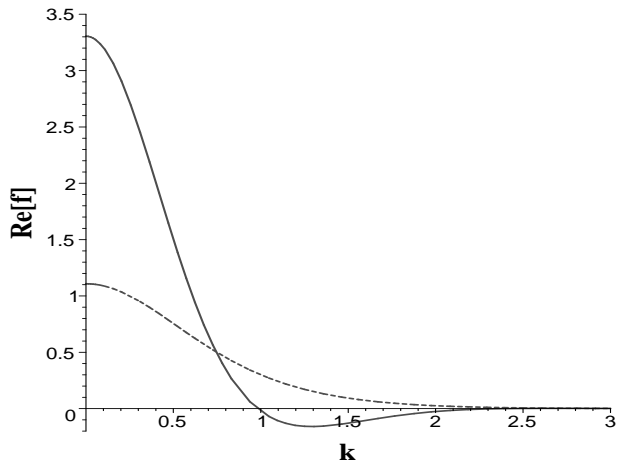}}
\caption{The real part of $f_{k}(\chi)$ as a function of $k$. The
solid line is for $\chi=0.1$ and the dashed line for $\chi=0.9$.}
\label{fig:1}
\end{figure}
\newpage
\begin{figure}
\centerline{\includegraphics[width=8 cm,height=6 cm]{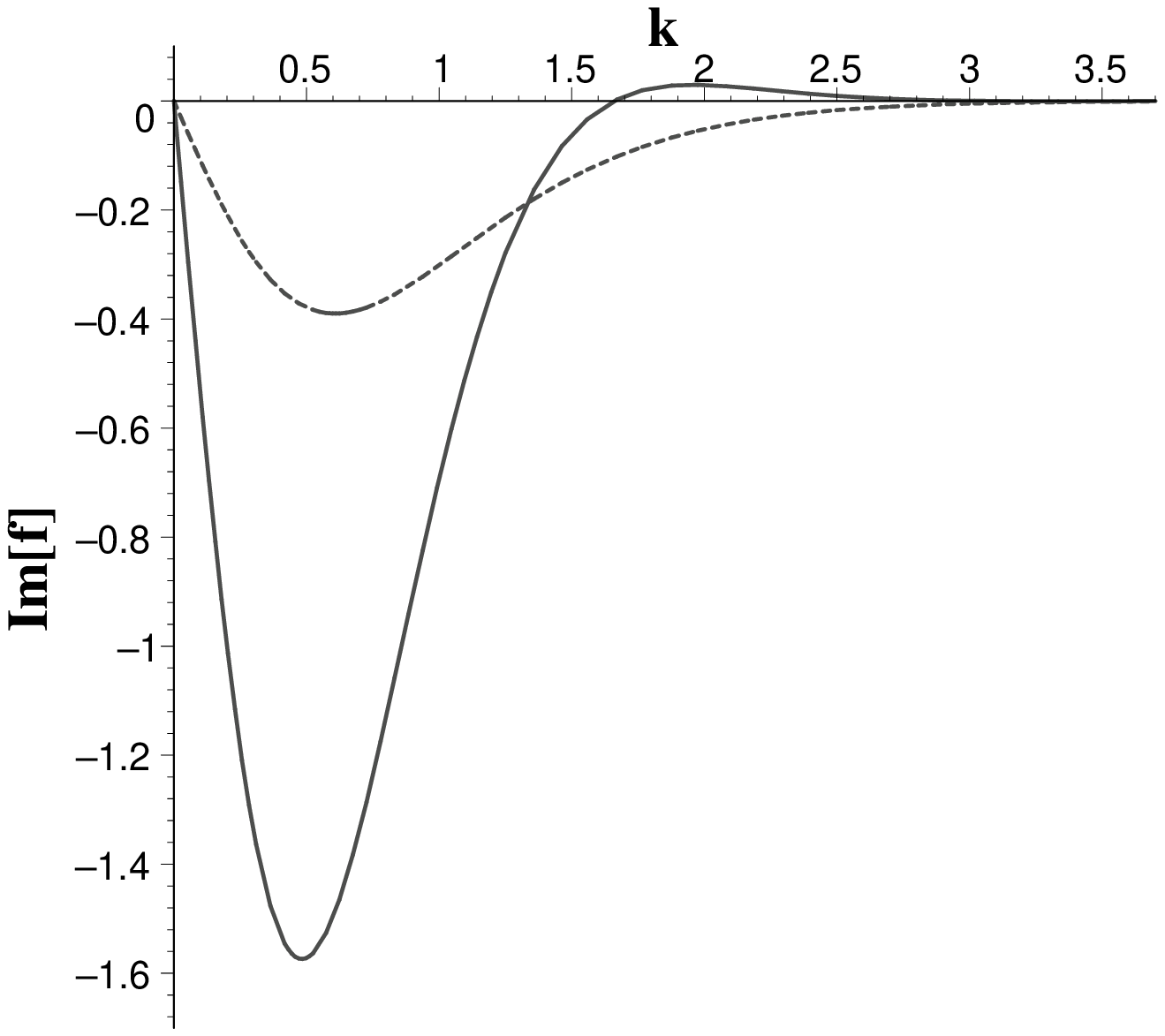}}
\caption{The imaginary part of $f_{k}(\chi)$ as a function of $k$.
The solid line is for $\chi=0.1$ and the dashed line for
$\chi=0.9$.}
\label{fig:2}
\end{figure}
\begin{figure}
\centerline{\includegraphics[width=8 cm,height=6 cm]{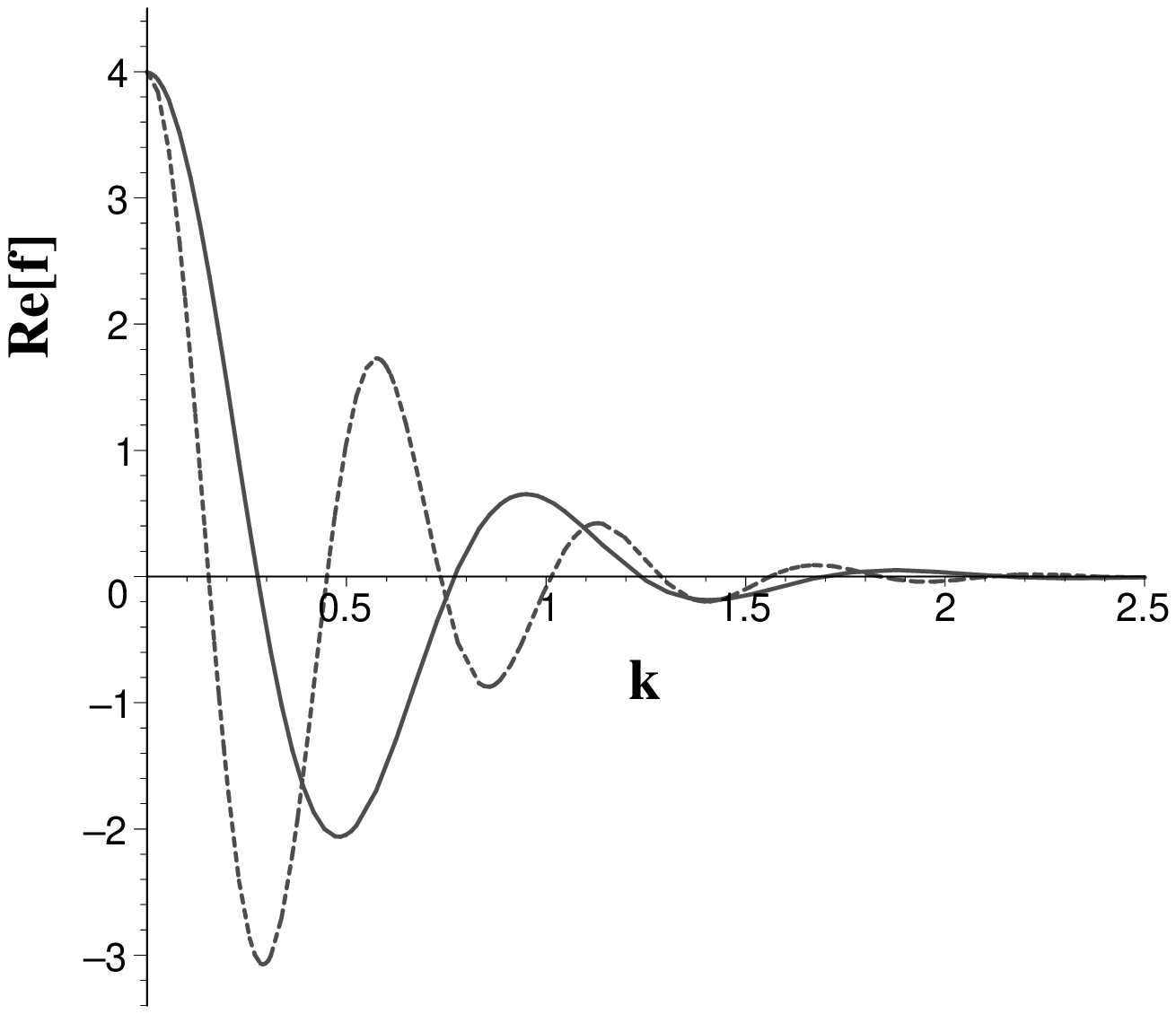}}
\caption{The real part of $f_{k}(\chi)$ as a function of $k$. The
solid line is for $\chi=0.001$ and the dashed line for
$\chi=0.00001$.}
\label{fig:3}
\end{figure}

\begin{figure}
\centerline{\includegraphics[width=8 cm,height=6 cm]{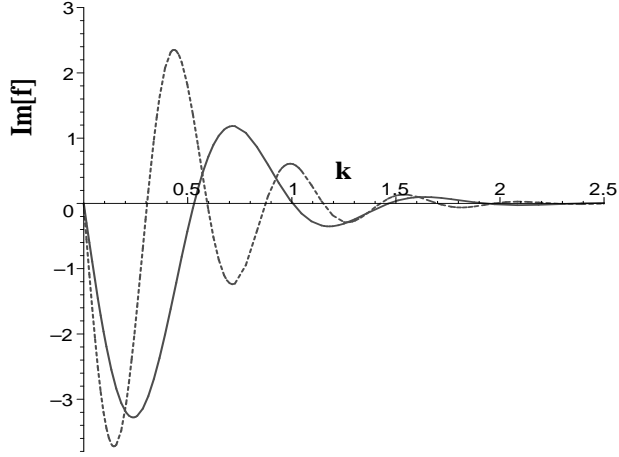}}
\caption{The imaginary part of $f_{k}(\chi)$ as a function of $k$.
The solid line is for $\chi=0.001$ and the dashed line for
$\chi=0.00001$.}
\label{fig:4}
\end{figure}

\newpage
 From our graphs we observe that both the real part and
the imaginary part of the function $f_{k}\left(\frac{p
}{p\,'}\right)$, are finite in origin and very convergent for
large values of $k$ (see Figs.(\ref{fig:1})-(\ref{fig:2})).
As the ratio of
the momenta $\frac{p}{p\,'}$ takes small values, we observe that
these functions become oscillatory (see
Figs.(\ref{fig:3})-(\ref{fig:4})). This oscillatory behavior is the
result of the behavior of Gauss hypergeometric functions as their
algebraic argument $1-\left(\frac{p}{p\,'}\right)^2$ approaches
to one (or $\frac{p}{p\,'}$ approaches zero), combined with the
oscillatory factors $\left(\frac{p}{p\,'}\right)\,^{1-ik}$. Squaring the amplitude and
summing after final helicities $\lambda,\lambda'$, we obtain the probability of pair production.

We find that the probability of fermion pair production in an electric field is:
\begin{eqnarray}\label{pr}
&&\mathcal{P}_{e^-e^+}=\frac{1}{2}\sum_{\lambda\lambda'}|\mathcal{A}_{e^-e^+}|\,^{2}\nonumber\\
&&=\frac{1}{2}\sum_{\lambda\lambda'}\frac{e^{4} }{(64)^2\pi^2
|\vec{p}+\vec{p}\,'|^{2}}
|\xi^{+}_{\lambda}(\vec{p}\,)(\vec{\sigma}\cdot\vec{n})\eta_{\lambda'}(\vec{p}\,\,')|^2
\{p\,'^{-4}\theta(p\,'-p)\left[2\left|f_{k}\left(\frac{p}{p\,'}\right)\right|^2\right.\nonumber\\
&&\left.-sgn(\lambda\lambda\,')\left(f^2_{k}\left(\frac{p}{p\,'}\right)+
f^{*\,2}_{k}\left(\frac{p}{p\,'}\right)\right)\right]+
p^{-4}\theta(p-p\,')\left[2\left|f_{k}\left(\frac{p\,'}{p}\right)\right|^2\right.\nonumber\\
&&\left.-sgn(\lambda\lambda\,')\left(f^2_{k}\left(\frac{p\,'}{p}\right)
+f^{*\,2}_{k}\left(\frac{p\,'}{p}\right)\right)\right]\}.
\label{prob}
\end{eqnarray}

\begin{figure}
\centerline{\includegraphics[width=8 cm,height=6 cm]{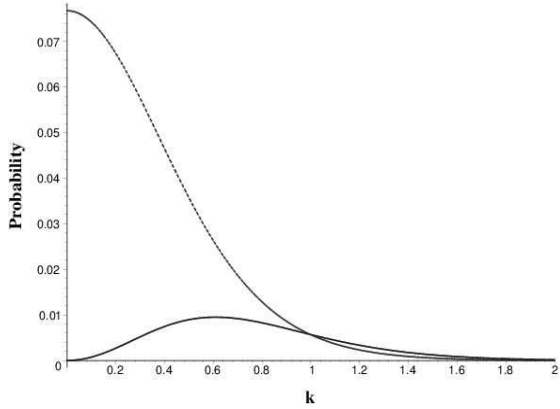}}
\caption{$\mathcal{P}_{e^-e^+}$ as a function of $k$ for $\chi=0.9$. The dashed line represents the case of helicity conservation and the solid line represents the case when helicity is not conserved.}
\label{fig:5}
\end{figure}

\begin{figure}
\centerline{\includegraphics[width=8 cm,height=6 cm]{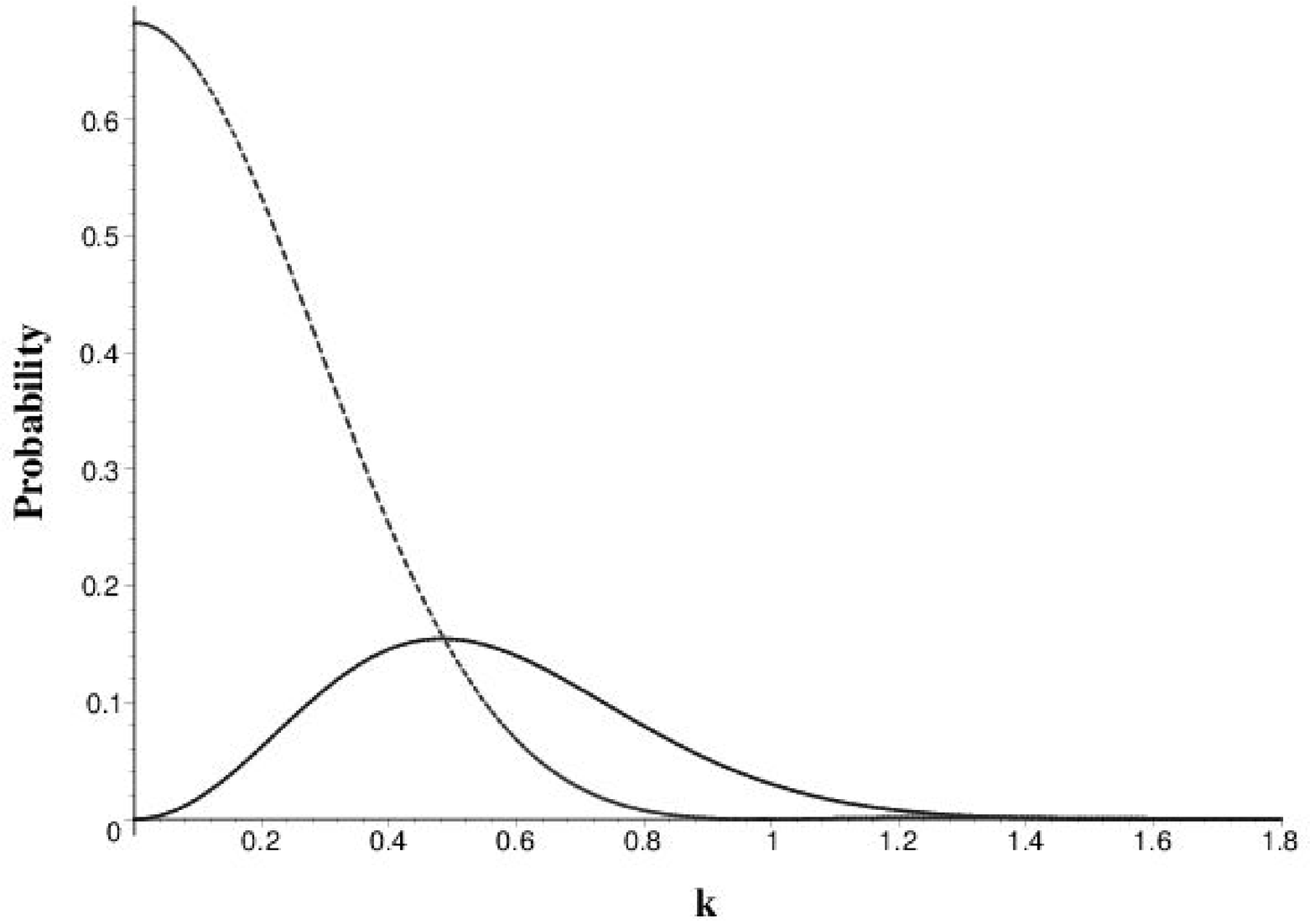}}
\caption{$\mathcal{P}_{e^-e^+}$ as a function of $k$ for $\chi=0.1$. The dashed line represents the case of helicity conservation and the solid line represents the case when helicity is not conserved.}
\label{fig:6}
\end{figure}

\begin{figure}
\centerline{\includegraphics[width=8 cm,height=6 cm]{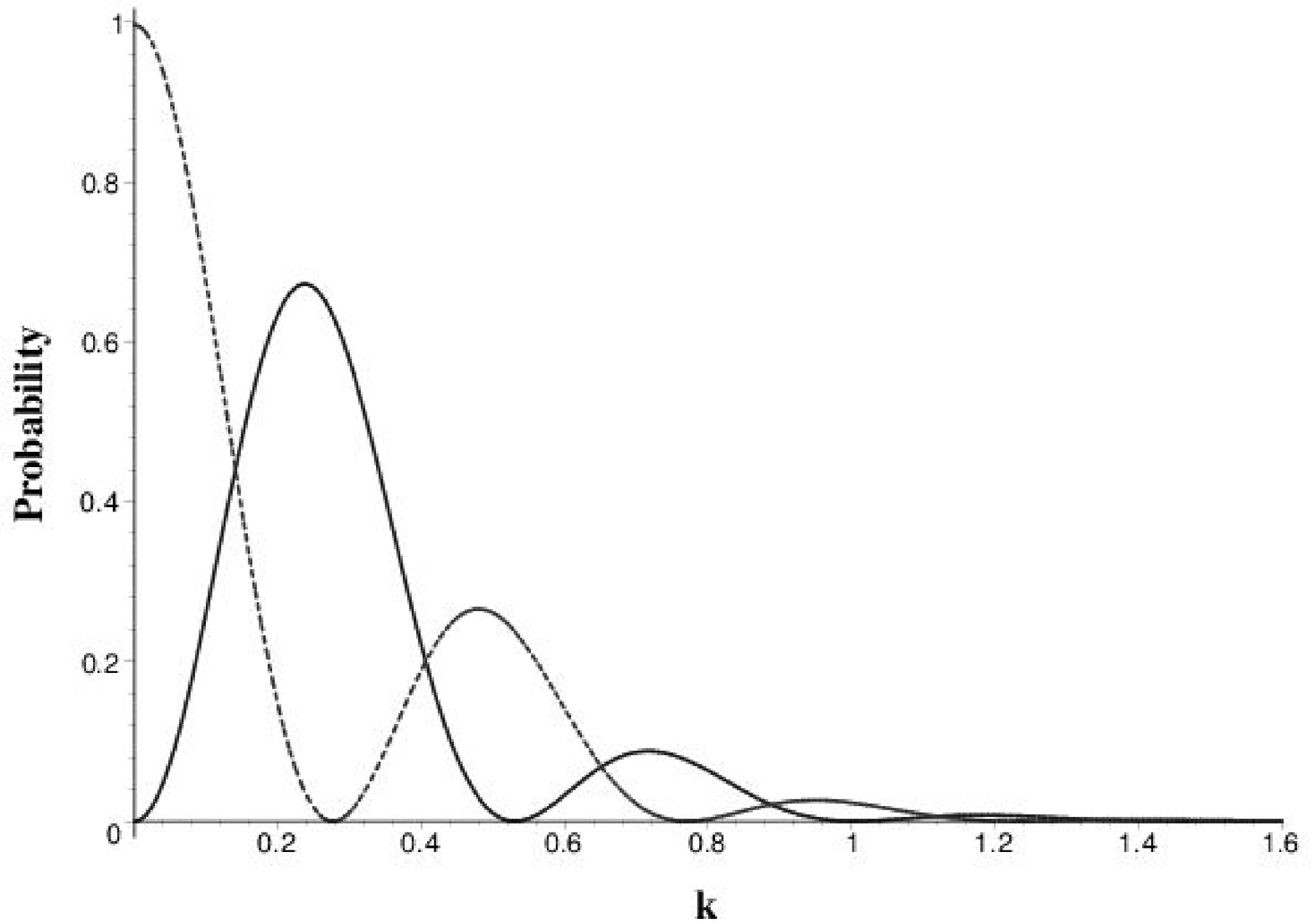}}
\caption{$\mathcal{P}_{e^-e^+}$ as a function of $k$ for $\chi=0.001$. The dashed line represents the case of helicity conservation and the solid line represents the case when helicity is not conserved.}
\label{fig:7}
\end{figure}

\begin{figure}
\centerline{\includegraphics[width=8 cm,height=6 cm]{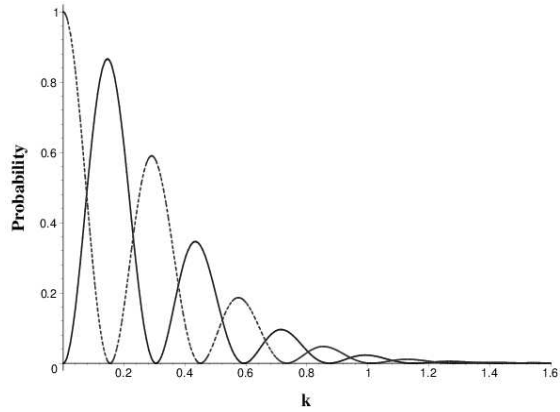}}
\caption{$\mathcal{P}_{e^-e^+}$ as a function of $k$ for $\chi=0.00001$. The dashed line represents the case of helicity conservation and the solid line represents the case when helicity is not conserved.}
\label{fig:8}
\end{figure}

The total probability is obtained integrating equation (\ref{prob})
after the final momenta $p,p\,'$: $\mathcal{P}^{tot}_{e^-e^+}=\int
\mathcal{P}_{e^-e^+}\,\frac{d^3p}{(2\pi)^{3}}\frac{d^3p\,'}{(2\pi)^{3}}$.
From the general formula for probability (\ref{prob}), we observe
that in the process of pair production by an electric field the
helicity is conserved when $\lambda=-\lambda\,'$ and the law
of helicity conservation is broken when $\lambda=\lambda\,'$.
We must specify that the non-conservation of helicity is driven by the fermion
mass and not by the de Sitter geometry \cite{26} .
Plotting our probability (\ref{pr}) as function of parameter $k$, we obtain the results from
Figs. (\ref{fig:5})-(\ref{fig:8}).

\newpage
From Figs.(\ref{fig:5})-(\ref{fig:8}) we observe that  the electric
field of the type considered here can produce pair of fermions
only in the early universe, when the expansion factor was sensibly
larger that the particle mass ($\omega > m$). This result confirms the predictions
from \cite{8,9,10} , were it was proven using a WKB approach, that the rate of pair production was
important only in the early universe. As $\chi\rightarrow 0$, the probability becomes oscillatory in
both conserving and non-conserving helicity cases. It is also
important to observe that the probabilities for a helicity conserving/non-conserving processes becomes approximatively equal in the limit
$\chi\rightarrow 0$ (see Fig. (\ref{fig:8})). Also from Figs. (\ref{fig:5})-(\ref{fig:8}),
we observe that it is more likely to produce pairs of fermions, which have the ratio of the momenta $\chi$ very small.
To conclude, the presence of an electric field in this geometry will favour the processes in which for example ,the electron
momenta will be small comparatively to the positron momenta, with the specification that large momenta don't
refer here, to the relativistic case. As $\chi$ is close to unity, we observe that the probability of pair production become sensibly
smaller (see Fig. (\ref{fig:5})), than in the case of small $\chi$ (see Figs. (\ref{fig:7}),(\ref{fig:8})).
We also observe that our probability is proportional with a factor $|p+p\,'|^2$, from which it is clear that in
the limit of large momenta our amplitude/probability
will vanish. From here we conclude that only fermions with small momenta were produced in the presence of an electric field,
due to the early expansion of the universe. 

The results from Figs. (\ref{fig:5})-(\ref{fig:8}), also show that the probability of pair production in electric field of a charge id sensibly larger that the probability of pair production in Coulomb field studied in \cite{26}. Also the present study show that there is a preferential direction of motion for the produced fermions which is the direction of the intensity of electric field , as we will see below.

Let us do a more detailed analysis in the helicity space.
We use helicity bispinors
$\xi_{\lambda}(\vec{p}\,)\,,\eta_{\lambda'}(\vec{p}\,\,')$, for
$\lambda=\lambda\,'$ and $\lambda=-\lambda\,'$, and restrict the
analysis only to the square of our amplitude. In an orthogonal
local frame $\{\vec{e}_i\}$, we take the electron and positron
momenta in the plane $(1,2)$, denoting their spherical coordinates
as $\vec{p}=(p,\alpha,\beta)$ ,
${\vec{p}\,}\,'=(p\,',\gamma,\delta)$ and we consider that the
unit vector which gives the direction and orientation of the
electric field, has the spherical coordinates
$\vec{n}=(1,\theta,\varphi)$ , where $\alpha,\, \gamma, \, \theta
\in(0,\pi);\,\beta,\,\delta,\, \varphi\in(0,2\pi)$.

Corresponding to the two cases of conserving/nonconserving
helicity in the pair production process by an electric field, the bispinor product
$\xi^{+}_{\lambda}(\vec{p}\,)(\vec{\sigma}\cdot\vec{n})\eta_{\lambda'}(\vec{p}\,\,')$ can be calculated.
In the case of helicity conservation this gives:
\begin{eqnarray}\label{hn}
\xi^{+}_{\frac{1}{2}}(\vec{p}\,)(\vec{\sigma}\cdot\vec{n})\eta_{-\frac{1}{2}}(\vec{p}\,\,')=\cos\left(\frac{\alpha}{2}\right)\cos\left(\frac{\gamma}{2}\right)\cos(\theta)
+e^{i(\varphi-\beta)}\sin\left(\frac{\alpha}{2}\right)
\cos\left(\frac{\gamma}{2}\right)\sin(\theta)\nonumber\\
+e^{i(\delta-\varphi)}\cos\left(\frac{\alpha}{2}\right)
\sin\left(\frac{\gamma}{2}\right)\sin(\theta)-e^{i(\delta-\beta)}\sin\left(\frac{\alpha}{2}\right)
\sin\left(\frac{\gamma}{2}\right)\cos(\theta),\nonumber\\
\xi^{+}_{-\frac{1}{2}}(\vec{p}\,)(\vec{\sigma}\cdot\vec{n})\eta_{\frac{1}{2}}(\vec{p}\,\,')=\cos\left(\frac{\alpha}{2}\right)\cos\left(\frac{\gamma}{2}\right)\cos(\theta)
+e^{-i(\varphi+\beta)}\sin\left(\frac{\alpha}{2}\right)
\cos\left(\frac{\gamma}{2}\right)\sin(\theta)\nonumber\\
+e^{i(\varphi-\delta)}\cos\left(\frac{\alpha}{2}\right)
\sin\left(\frac{\gamma}{2}\right)\sin(\theta)-e^{-i(\delta+\beta)}\sin\left(\frac{\alpha}{2}\right)
\sin\left(\frac{\gamma}{2}\right)\cos(\theta).
\nonumber\\
\end{eqnarray}
For the helicity nonconserving case the result is:
\begin{eqnarray}\label{hc}
\xi^{+}_{\frac{1}{2}}(\vec{p}\,)(\vec{\sigma}\cdot\vec{n})\eta_{\frac{1}{2}}(\vec{p}\,\,')=e^{-i\delta}\cos\left(\frac{\alpha}{2}\right)\sin\left(\frac{\gamma}{2}\right)\cos(\theta)
+e^{i(\varphi-\beta-\delta)}\sin\left(\frac{\alpha}{2}\right)
\sin\left(\frac{\gamma}{2}\right)\sin(\theta)\nonumber\\
+e^{-i\beta}\sin\left(\frac{\alpha}{2}\right)
\cos\left(\frac{\gamma}{2}\right)\cos(\theta)-e^{-i\varphi}\cos\left(\frac{\alpha}{2}\right)
\cos\left(\frac{\gamma}{2}\right)\sin(\theta),\nonumber\\
\xi^{+}_{-\frac{1}{2}}(\vec{p}\,)(\vec{\sigma}\cdot\vec{n})\eta_{-\frac{1}{2}}(\vec{p}\,\,')=-e^{i\delta}\cos\left(\frac{\alpha}{2}\right)\sin\left(\frac{\gamma}{2}\right)\cos(\theta)
-e^{i(\delta-\varphi-\beta)}\sin\left(\frac{\alpha}{2}\right)
\sin\left(\frac{\gamma}{2}\right)\sin(\theta)\nonumber\\
-e^{-i\beta}\sin\left(\frac{\alpha}{2}\right)
\cos\left(\frac{\gamma}{2}\right)\cos(\theta)+e^{i\varphi}\cos\left(\frac{\alpha}{2}\right)
\cos\left(\frac{\gamma}{2}\right)\sin(\theta).
\nonumber\\
\end{eqnarray}
Taking now $\beta=\pi\,,\delta=\varphi=0$ and squaring the
bispinor product we obtain:
\begin{eqnarray}\label{bis}
|\xi^{+}_{\lambda}(\vec{p}\,)(\vec{\sigma}\cdot\vec{n})\eta_{\lambda'}(\vec{p}\,\,')|^2=\left\{
\begin{array}{cll}
|\cos(\theta)\cos\left(\frac{\alpha-\gamma}{2}\right)-\sin(\theta)\sin\left(\frac{\alpha-\gamma}{2}\right)|^2\,,\,&{\rm for}&\lambda=-\lambda'\\
|\cos(\theta)\sin\left(\frac{\alpha-\gamma}{2}\right)+\sin(\theta)\cos\left(\frac{\alpha-\gamma}{2}\right)|^2,\,&{\rm
for}&\lambda=\lambda'.
\end{array}\right.
\nonumber\\
\end{eqnarray}
From equation (\ref{bis}) if we set $\alpha=\pi, \gamma=
\theta=0$, corresponding to the situation when the
electron-positron pair is generated on the direction of electric
field, but they move in the opposite senses, we observe that the
probability in the helicity conserving case is zero, while the
probability of a process where helicity is not conserved becomes
maxim. From here we can conclude that only the production processes that don't conserve the
helicity could produce pairs that move in opposite senses, increasing in
this way the chances of separation between matter and anti-matter.
Now if we set $\alpha=\gamma=\theta=0$, corresponding to the
situation when the electron-positron pair is emitted on the
direction of the electric field, but they move in the same sense, we
observe that the probability in the helicity conserving case
becomes maxim and is zero in the helicity non-conserving case. In
this case, it is more likely that the pair will annihilate than
become separate. Also from
(\ref{hn}),(\ref{hc}),(\ref{bis}), we observe that the electron-positron pair could be emitted
in other directions, which make various angles with the direction
of the electric field, but with smaller probabilities.

To conclude, in both helicity conserving/non-conserving cases,
the most probable transitions are those where the pairs are emitted on the direction of the electric field.

\section{Number of fermions}

Let us translate our perturbational result in terms of density number of produced particles.
In the vast majority of the papers related to the particle production, the outcome of the calculations
is the density number of created particles. So our reader could ask how we will compute the density number, taking into account that we use a perturbative approach.

First let us comment on the previous results related to the density number of created particles. The most important result in this direction was obtained by L.Parker using a WKB approach \cite{8,9,10}. This result shows for the first time that the density number of created particles was important only in the early universe, which corresponds to the inflation phase.
However there are results concerning the density number of created particles obtained using the method of Bogoliubov transformations. As was shown by Fulling \cite{30}, there are many cases of interest where this method gives divergent results. This happens because the density number of created particles is not a function of the final momenta and as a result of that, the integration after the final momenta will give a quantity which is linearly divergent. This important observation is valid for the results presented in \cite{19,20}, where an integration after final momenta gives an infinite number of particles. So we can not even speak about a regularization of the integral simply because we do not have any dependence on final momenta \cite{19,20}. In addition, when we compute the number of created particles in an expanding universe, this will be in general a time dependent quantity. This is the result of the fact the the volume will expand in time. For example, in de Sitter case discussed here, the physical spatial coordinates are $x_{i}^{ph}=x_{i}\,e^{\omega t}$, and from here we observe that volume will expand as $V_{ph}=V\,e^{3\omega t}$ . The above mentioned problems seem to receive little attention despite their crucial importance.  In \cite{29} the density number of created particles was computed in a Robertson-Walker metric, and this is one of the few papers which takes into account the effect of the field interactions upon particle production processes. Moreover in \cite{29}, the correct density number of created particles was computed taking into account that the volume is a function of the expansion factor.

Here we give the main steps for computing the number of created fermions in our case. Squaring the amplitude and summing after the final helicities $\lambda,\,\lambda'$ we obtain the probability of pair production (\ref{pr}). Then following the result from \cite{29}, the ratio probability/volume, is the density  number of fermions :
\begin{equation}\label{numd}
\mathcal{N}=\frac{\mathcal{P}}{V_{ph}}=\frac{1}{2}\sum_{\lambda\lambda'}\frac{|\mathcal{A}_{e^-e^+}|\,^{2}}{V\,e^{3\omega t}}.
\end{equation}
This quantity can be computed now explicitly using the equation for probability (\ref{pr}).
The total number of produced fermions can be obtained integrating the density number above, after the final momenta $p,\,p'$:
\begin{equation}\label{num}
\mathcal{N}^{tot}=\frac{\mathcal{P}^{tot}}{V_{ph}}=\int\frac{\mathcal{P}}{V\,e^{3\omega t}}\frac{d^3p}{(2\pi)^{3}}\frac{d^3p\,'}{(2\pi)^{3}}.
\end{equation}
Here a few comments are in order. First our density number of produced fermions defined in (\ref{numd}), is dependent on the final momenta $p,\,p\,'$. However the integrals that result are not known in mathematics, these being a combination of two hypergeometric Gauss functions and a fraction of the type $\frac{p^{\nu}p'^{\mu}}{|\vec{p}+\vec{p}\,'|^{2}}$, where $\nu,\,\mu$ are complex numbers. Despite of these difficulties we will further give a way to calculate the total number of fermions.

First from our graphs for probability Figs. (\ref{fig:5})-(\ref{fig:8}), we observe that there are larger probabilities to produce pairs of fermions which have the ratio of the momenta very small $\chi<<1$. This is the case when  for example the positron momenta is large comparatively with the electron momenta ($p\,'>> p$).
In this case the ratio $\left(\frac{p}{p\,'}\right)^\alpha\simeq0$, when $\alpha$ is a real number and the hypergeometric function become $_{2}F_{1}\left(a,b;c;0\right)=1$. Then the final expression of the function $f_{k}$ for $p\,'>>p$, is expressed only in purely imaginary powers:
\begin{eqnarray}\label{fcf}
f_{k}\left(\frac{p}{p\,'}\right)\simeq\frac{4}{\sqrt{\pi}}
\left(\frac{p}{p\,'}\right)^{ik}
\Gamma\left(\frac{1}{2}-ik\right)\Gamma(1+ik).
\nonumber\\
\end{eqnarray}
The total probability will be calculated for the helicity nonconserving case and we will consider the most probable transitions which produce pairs that move on the direction of electric field. In the previous section we see that only the helicity nonconserving processes could produce pairs that are separable. This is equivalently to take $\vec{p}\vec{p}\,'=|\vec{p}\,||\vec{p}\,'|\cos(\alpha+\gamma)=-|\vec{p}\,||\vec{p}\,'|$, corresponding to an angle between momenta vectors equal with $\pi$ (helicity nonconservation) and from here $|\vec{p}+\vec{p}\,'|^{2}\simeq p\,'^2-2pp\,'$. For $\lambda=\lambda'$ the summation after helicity bispinors become:
\begin{eqnarray}\label{bi}
\sum_{\lambda}|\xi^{+}_{\lambda}(\vec{p}\,)(\vec{\sigma}\cdot\vec{n})\eta_{\lambda}(\vec{p}\,\,')|^2&=&|\xi^{+}_{\frac{1}{2}}(\vec{p}\,)(\vec{\sigma}\cdot\vec{n})\eta_{\frac{1}{2}}(\vec{p}\,\,')|^2+
|\xi^{+}_{-\frac{1}{2}}(\vec{p}\,)(\vec{\sigma}\cdot\vec{n})\eta_{-\frac{1}{2}}(\vec{p}\,\,')|^2
\end{eqnarray}
and if we consider that the most probable transitions in the helicity nonconserving case are obtained when $\alpha=\pi,\,\gamma=\theta=0$
, corresponding to the situation when electron and positron move in opposite senses, the sum after helicity bispinors (\ref{bi}), simplify to a numeric factor.  For the momenta modulus integrals we consider that $p\,'\in (0,\infty)$ and we restrict the limits of integration for $p$ to $p \in (\mu,\Lambda)$. The cutoff for the upper limit of integration in the electron momenta $p$ is in accord with our suppositions that $p<<p\,'$ . We also take the lower limit of integration for $p$ to be $\mu$.

Using (\ref{fcf}) we compute the quantity that defines our probability.
For $p\,'>>p$, we neglect the terms $\left(\frac{p}{p\,'}\right)^\alpha\simeq 0$,(with $\alpha$ a real number )  and we take only terms with purely imaginary powers, the result being:
\begin{eqnarray}\label{apr}
&&2\left|f_{k}\left(\frac{p}{p\,'}\right)\right|^2-f^2_{k}\left(\frac{p}{p\,'}\right)-f^{*\,2}_{k}\left(\frac{p}{p\,'}\right)\simeq\frac{16}{\pi}
\left[2\left|\Gamma\left(\frac{1}{2}-ik\right)\right|^2|\Gamma(1+ik)|^2\right.\nonumber\\
&&\left.-\Gamma^2\left(\frac{1}{2}-ik\right)\Gamma^2(1+ik)\left(\frac{p}{p\,'}\right)^{2ik}-\Gamma^2\left(\frac{1}{2}+ik\right)\Gamma^2(1-ik)
\left(\frac{p}{p\,'}\right)^{-2ik}\right].
\end{eqnarray}
We will note $A=\Gamma^2\left(\frac{1}{2}-ik\right)\Gamma^2(1+ik)\,\,,B=\Gamma^2\left(\frac{1}{2}+ik\right)\Gamma^2(1-ik)$.
Then the  total number of produced fermions in volume unit for the case when electron-positron pair move in opposite senses (helicity nonconserving case), but on the direction of the electric field vector, for  $p\,'>>p$ is:
\begin{eqnarray}
&&\mathcal{N}^{tot}=\frac{e^{4}}{256\pi^3\,e^{3\omega t}}
\int \frac{d\Omega_{p}}{(2\pi)^3}\int \frac{d\Omega_{p\,'}}{(2\pi)^3}\nonumber\\
&&\times\Re \biggl\{ \int_{\mu}^{\Lambda}
\left[\int_{0}^{\infty} \frac{dp\,'}{p\,'^{2}(p\,'^{2}- 2pp\,')}\left(-A\left(\frac{p}{p\,'}\right)^{2ik}-B\left(\frac{p}{p\,'}\right)^{-2ik}\right)\right]p^2 dp\biggl\},
\end{eqnarray}
where $\Re\{..\}$ is the real part of our result. This is because we integrate the ratio of the momenta at imaginary powers and the final result will be an imaginary number. The second observation here is that we neglected the first term in (\ref{apr}) due to the fact that will be proportional when replaced in our probability with $\left(\frac{p}{p\,'}\right)^{2}$, which is negligible.
The integrals that help us to establish the expression for total number of fermions are given in Appendix.

Before proceed in our calculations for the number of fermions we need to make a few clarifications. We observe that the integrand will contain momenta at imaginary powers and when integrated the final result will be an imaginary number. This is the result of the fact that we equal with unity the hypergeometric functions which contain imaginary arguments. For these reasons we must interpret the total number of fermions as the real part of our result.
Then the final result for total number of fermions in volume unit will be:
\begin{eqnarray}
&&\mathcal{N}^{tot}=\frac{\alpha^{2}}{512\pi^4 \sinh(2\pi k)}(\ln\Lambda-\ln\mu)e^{-3\omega t}\nonumber\\&&\times\Re\biggl\{i\Gamma^2\left(\frac{1}{2}-ik\right)\Gamma^2(1+ik)e^{-2\pi k}2^{-2ik}-i\Gamma^2\left(\frac{1}{2}+ik\right)\Gamma^2(1-ik)e^{2\pi k}2^{2ik}\biggl\}\
\nonumber\\&&=\frac{\alpha^{2}e^{-3\omega t}}{512\pi^4}g_{k}(\mu,\Lambda),
\end{eqnarray}
where we note
\begin{eqnarray}
g_{k}(\mu,\Lambda)&=&\Re\biggl\{i\Gamma^2\left(\frac{1}{2}-ik\right)\Gamma^2(1+ik)e^{-2\pi k}2^{-2ik}-i\Gamma^2\left(\frac{1}{2}+ik\right)\Gamma^2(1-ik)e^{2\pi k}2^{2ik}\biggl\}\nonumber\\
&& \times\frac{(\ln\Lambda-\ln\mu)}{\sinh(2\pi k)}.
\end{eqnarray}
We see that the total number of fermions is a time dependent quantity and decreases exponentially in time. This is natural since the expansion factor decreases also with time. From here we can conclude that the effect presented in this paper could be observable only in strong gravitational fields. Another interesting observation is that the total probability and total number of fermions contains both ultraviolet and infrared divergences that are logarithmical.

\begin{figure}
\centerline{\includegraphics[width=8 cm,height=6 cm]{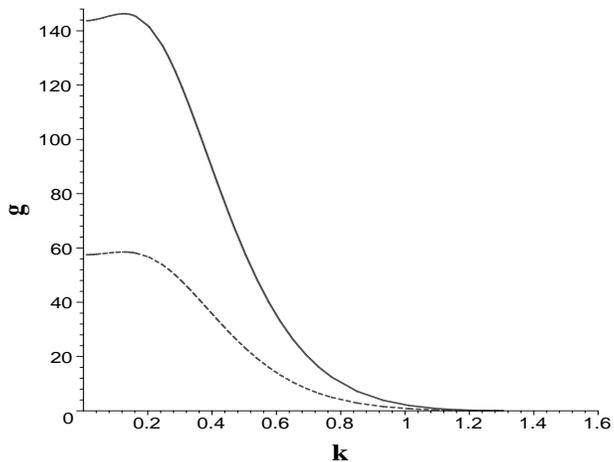}}
\caption{$g_{k}(\mu,\Lambda)$ as a function of $k$. The dashed line represents the case when $\mu=0.001,\,\Lambda=1000$ and the solid line represents the case when $\mu=0.000001,\,\Lambda=1000000000$.}
\label{fig:9}
\end{figure}

\begin{figure}
\centerline{\includegraphics[width=8 cm,height=6 cm]{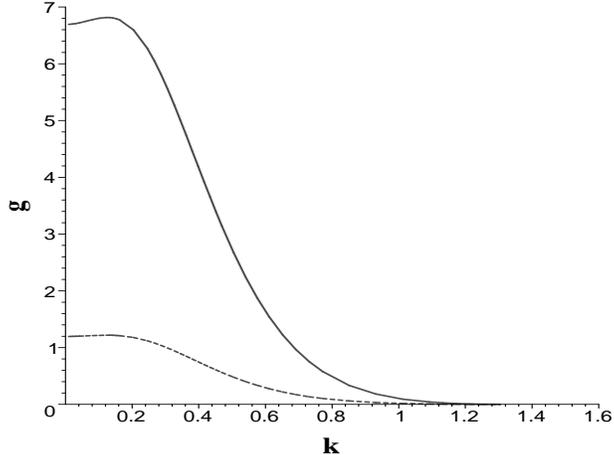}}
\caption{$g_{k}(\mu,\Lambda)$ as a function of $k$. The dashed line represents the case when $\mu=3,\,\Lambda=4$ and the solid line represents the case when $\mu=2,\,\Lambda=10$.}
\label{fig:10}
\end{figure}

Plotting the function $g_{k}(\mu,\Lambda)$ in terms of parameter $k$ for different values of $\mu,\, \Lambda$ we obtain the results presented in Figs. (\ref{fig:9})-(\ref{fig:10}). From these graphs we observe that the number of fermions produced by the electric field considered here was important only in the early universe. When the upper limit in momentum integration is large comparatively with the lower limit $\Lambda\gg \mu$, one can observe that the number of produced fermions is sensibly bigger than in the case when $\Lambda\simeq \mu$.

\newpage
\section{Conclusions}
We compute the amplitude/probability for pair production by the electric field of a charge in the de Sitter expanding universe. The probability of pair production is nonvanishing only in the early universe when the expansion factor was large. From our calculations we recover the Minkowski limit, where the amplitude and probability for this process vanish. We found that that the most probable transitions are those in which the pairs are emitted on the direction of the electric field.

We also translate our perturbational calculations in terms of density number of fermions. A method to compute the total number of fermions in the case when the ratio $\frac{p}{p\,'}$ is close to zero, was proposed. The total number of produced fermions was obtained and we prove that is nonvanishing only in the early universe.

\section{Appendix}
Let us introduce unit normalized helicity spinors \cite{6} for an
arbitrary momentum vector $\vec{p}$ : $\xi_{\lambda}(\vec{p}\,)$
and $\eta_{\lambda}(\vec{p}\,)= i\sigma_2
[\xi_{\lambda}(\vec{p}\,)]^{*}$
\begin{equation}\label{pa}
\vec{\sigma}\vec{p}\,\xi_{\lambda}(\vec{p}\,)=2p\lambda\xi_{\lambda}(\vec{p}\,)
\end{equation}
with $\lambda=\pm1/2$ and where $\vec{\sigma}$ are the Pauli
matrices and $p=\mid\vec{p}\mid$. The particle spinors have the
form
\begin{equation}\label{xi}
\xi_{\frac{1}{2}}(\vec{p}\,)=\sqrt{\frac{p_3+p}{2 p}}\left(
\begin{array}{c}
1\\
\frac{p_1+ip_2}{p_3+p}
\end{array} \right)\,,\quad
\xi_{-\frac{1}{2}}(\vec{p}\,)=\sqrt{\frac{p_3+p}{2 p}}\left(
\begin{array}{c}
\frac{-p_1+ip_2}{p_3+p}\\
1
\end{array} \right)\,.
\end{equation}

Then, the positive/negative frequency modes of momentum $\vec{p}$
and helicity $\lambda$ derived in \cite{4} , assuming
 gamma matrices in Dirac representation, are \cite{1,4,5} :
\begin{eqnarray}\label{sol}
U_{\vec{p},\lambda}(t,\vec{x}\,)=\frac{\sqrt{\pi
p/\omega}}{(2\pi)^{3/2}}\left (\begin{array}{c}
\frac{1}{2}e^{\pi k/2}H^{(1)}_{\nu_{-}}(\frac{p}{\omega} e^{-\omega t})\xi_{\lambda}(\vec{p}\,)\\
\lambda e^{-\pi k/2}H^{(1)}_{\nu_{+}}(\frac{p}{\omega} e^{-\omega
t})\xi_{\lambda}(\vec{p}\,)
\end{array}\right)e^{i\vec{p}\cdot\vec{x}-2\omega t},
\end{eqnarray}
where $H^{(1)}_{\nu}(z)$ are Hankel functions of
the first kind and
$k=\frac{m}{\omega},\nu_{\pm}=\frac{1}{2}\pm ik$. The negative frequency modes will be obtained via charge conjugation operation $V_{\vec{p},\lambda}(t,\vec{x}\,)=i\gamma^2 \gamma^0 (\bar U_{\vec{p},\lambda}(t,\vec{x}\,))^{T}$.

For solving our integrals with Hankel functions we use their
relations with Macdonald functions \cite{11,12,23} :
\begin{equation}
H^{(1,2)}_{\nu}(z)=\mp \left(\frac{2i}{\pi}\right)e^{\mp
i\pi\nu/2}K_{\nu}(\mp iz).
\end{equation}
In this way we arrive at the integrals of the type \cite{23} :
\begin{eqnarray}\label{a0}
&&\int_0^{\infty} dz
z^{-\lambda}K_{\mu}(az)K_{\nu}(bz)=\frac{2^{-2-\lambda}a^{-\nu+\lambda-1}b\,^{\nu}}{\Gamma(1-\lambda)}\Gamma\left(\frac{1-\lambda+\mu+\nu}{2}\right)\Gamma\left(\frac{1-\lambda-\mu+\nu}{2}\right)\nonumber\\
&&\times\Gamma\left(\frac{1-\lambda+\mu-\nu}{2}\right)\Gamma\left(\frac{1-\lambda-\mu-\nu}{2}\right)
\,_{2}F_{1}\left(\frac{1-\lambda+\mu+\nu}{2},\frac{1-\lambda-\mu+\nu}{2};1-\lambda;1-\frac{b^2}{a^2}\right),\nonumber\\
&&Re(a+b)>0\,,Re(\lambda)<1-|Re(\mu)|-|Re(\nu)|.
\end{eqnarray}
In our case $\lambda=-1$ and the second condition for convergence
is satisfied. We also observe that in our case $a,b$ are complex
and for solving our integrals we add to $a$ a small real part
$a\rightarrow a+\epsilon$, with $\epsilon>0$ and in the end we
take the limit $\epsilon\rightarrow 0$. This assure the
convergence of our integral and will correctly define the unit
step functions and $f_{k}$ functions.

For the computation of the total number of particles we use the next relation between hypergeometric functions:
\begin{eqnarray}\label{a1}
_{2}F_{1}(a,b\,;c;z)=\frac{\Gamma(c)\Gamma(c-a-b)}{\Gamma(c-a)\Gamma(c-b)}\,_{2}F_{1}(a,b\,;a+b-c+1;1-z)\nonumber\\
+(1-z)^{c-a-b}\,\frac{\Gamma(c)\Gamma(a+b-c)}{\Gamma(a)\Gamma(b)}\,_{2}F_{1}(c-a,c-b\,;c-a-b+1;1-z).
\end{eqnarray}
The integrals that helps us to calculate the total number of fermions are:
\begin{equation}
\int_{0}^{\infty}\frac{dp\,'p\,'^{-2\mp 2ik}}{(p\,'^{2}- 2pp\,')}=\mp\frac{i\pi2^{(\mp2ik)}}{8\,p^3\sinh(2\pi k)}\left(-\frac{1}{p}\right)^{\pm 2ik};\,\,\,\,\int_{\mu}^{\Lambda}\frac{dp}{p}=\ln(\Lambda)-\ln(\mu).
\end{equation}

\end{document}